\documentclass[12pt]{article}
\usepackage{epsfig}
\usepackage{axodraw}

\newcommand{\alphas}{\alpha_{\mathrm{s}}}

\newcommand{\xF}{x_{\mathrm{F}}}
\newcommand{\pt}{p_{\perp}}
\newcommand{\kt}{k_{\perp}}
\renewcommand{\b}{\mathrm{b}}
\renewcommand{\c}{\mathrm{c}}
\renewcommand{\d}{\mathrm{d}}
\newcommand{\e}{\rm{e}}

\newcommand{\g}{\mathrm{g}}

\newcommand{\p}{\mathrm{p}}
\newcommand{\q}{\mathrm{q}}
\renewcommand{\u}{\mathrm{u}}
\renewcommand{\d}{\mathrm{d}}
\newcommand{\s}{\mathrm{s}}
\newcommand{\D}{\mathrm{D}}

\newcommand{\Q}{\mathrm{Q}}
\newcommand{\V}{\mathrm{V}}

\newcommand{\bbar}{\overline{\mathrm{b}}}
\newcommand{\cbar}{\overline{\mathrm{c}}}
\newcommand{\dbar}{\overline{\mathrm{d}}}
\newcommand{\qbar}{\overline{\mathrm{q}}}
\newcommand{\sbar}{\overline{\mathrm{s}}}
\newcommand{\ubar}{\overline{\mathrm{u}}}
\newcommand{\Bbar}{\overline{\mathrm{B}}}
\newcommand{\Dbar}{\overline{\mathrm{D}}}
\newcommand{\Qbar}{\overline{\mathrm{Q}}}

\newcommand{\Py}{{\sc{Pythia}}}
\newcommand\hepph[1]{hep-ph/#1}
\newcommand\npb[3]{Nucl. Phys. {\bf B#1}, #3 (#2)}
\newcommand\ibid[3]{ibid. {\bf #1}, #3 (#2)}

{\end{list}}
\newcounter{enumct}

\newlength{\abstwidth}
\begin{document}

\sloppy

\pagestyle{empty}

\begin{flushright}
LU TP 00--10\\
LBNL-45275\\
March 2000
\end{flushright}
 
\vspace{\fill}

\begin{center}
{\LARGE \bf Bottom Production Asymmetries\\
at the LHC\footnote{To appear in the proceedings of the
``CERN 1999 Workshop on SM physics (and more) at the LHC''.}}\\[10mm]
{\Large E. Norrbin\footnote{emanuel@thep.lu.se}}\\[3mm]
{\it Department of Theoretical Physics,}\\
{\it Lund University, Lund, Sweden}\\[5mm]
{\large and}\\[5mm]
{\Large R. Vogt\footnote{vogt@lbl.gov}} \\[3mm]
{\it Physics Department,}\\
{\it University of California at Davis}\\
{\it and}\\
{\it Nuclear Science Division,}\\
{\it Lawrence Berkeley National Laboratory
\footnotetext{This work was supported in part by the Director, Office of Energy
Research, Division of Nuclear Physics of the Office of High Energy
and Nuclear Physics of the U. S. Department of Energy under Contract
Number DE-AC03-76SF00098.}}\\[3mm]
\end{center}

\vspace{\fill}
 
\begin{center}
{\bf Abstract}\\[2ex]
\begin{minipage}{\abstwidth}
We present results on bottom hadron production asymmetries at the LHC within both
the Lund string fragmentation model and the intrinsic bottom model. The main
aspects of the models are summarized and specific predictions for $\p\p$ collisions
at 14 TeV are given. Asymmetries are found to be very small at central
rapidities increasing to a few percent at forward rapidities. At very large
rapidities intrinsic production could dominate but this region is probably out
of reach of any experiment.
\end{minipage}
\end{center}
 
\vspace{\fill}

\clearpage
\pagestyle{plain}
\setcounter{page}{1}

\section{Introduction}
Sizeable leading particle asymmetries between e.g. $\D^-$ and $\D^+$ have been observed
in several fixed target experiments \cite{Asymobs}. It is of interest to
investigate to what extent these phenomena translate to bottom production and higher
energies. No previous experiment has observed asymmetries for bottom hadrons due to
limited statistics or other experimental obstacles.
Bottom asymmetries are in general expected to be smaller than for charm because of the
larger bottom mass, but there is no reason why they should be absent. In the fixed target
experiment HERA-B, bottom asymmetries could very well be large \cite{Norrbin} even at
central rapidities, but the conclusion of the present study is that asymmetries at the
LHC are likely to be small. In the following we study possible asymmetries between
$\mathrm{B}$ and $\overline{\mathrm{B}}$ hadrons at the LHC within the Lund string
fragmentation model \cite{Norrbin:PLB442} and the intrinsic heavy quark model
\cite{Brodsky:PLB93}.

In the string fragmentation model \cite{AGIS:PRP97},
the perturbatively produced heavy quarks are colour connected to
the beam remnants. This gives rise to beam-drag effects where the
heavy hadron can be produced at larger rapidities than the heavy quark.
The extreme case in this direction is the collapse of a small string,
containing a heavy quark and a light beam remnant valence quark of the
proton, into a single hadron. This gives rise to flavour correlations
which are observed as asymmetries. Thus, in the string model, there can be
coalescence between a perturbatively produced bottom quark
and a light quark in the beam remnant producing a leading bottom hadron.

There is also the possibility to have coalescence between the light valence quarks
and bottom quarks already present in the proton, because the wavefunction of the proton
can fluctuate into Fock configurations containing a $\b \bbar$ pair, such
as $|\u\u\d\b\bbar\rangle$.  In these states, two or more gluons are
attached to the bottom quarks, reducing the amplitude by ${\cal O}(\alphas^2)$
relative to parton fusion \cite{Vogt:NPB438}.  The longest-lived fluctuations in
states with invariant mass $M$ have a lifetime of ${\cal O}( 2 P_{\rm
lab}/M^2)$ in the target rest frame, where $P_{\rm lab}$ is the 
projectile momenta. Since
the comoving bottom and valence quarks have the same rapidity in these states,
the heavy quarks carry a large fraction of the projectile momentum and can thus
readily combine to produce bottom hadrons with large
longitudinal momenta. Such a mechanism can then dominate the
hadroproduction rate at large $\xF$. This is the underlying
assumption of the intrinsic heavy quark model \cite{Brodsky:PLB93},
in which the wave function fluctuations are initially
far off shell. However, they materialize as heavy hadrons when light spectator
quarks in the projectile Fock state interact with the target \cite{Brodsky:NPB369}.

In both models the coalescence probability is largest at small relative rapidity
and rather low transverse momentum where the invariant mass of the
$\Qbar\q$ system is small, enhancing the binding amplitude. One exception is
at very large $\pt$, where the collapse of a scattered valence quark with
a $\bbar$ quark from the parton shower is also possible, giving a further (small) source
of leading particle asymmetries in the string model.

\section{Lund String Fragmentation}
\label{Asym:Lundstring}
Before describing the Lund string fragmentation model, some words on the perturbative
heavy quark production mechanisms included in the Monte Carlo event generator \mbox{\Py}~\cite{Pythia:ref}
used in this study is in order. We study $\p\p$ events with one hard
interaction because events with no hard interaction are not
expected to produce heavy flavours and events with more than one hard interaction --- multiple
interactions ---  are beyond the scope of this initial study and presumably would
not influence the asymmetries. After the hard interaction
is generated, parton showers are added, both to the initial (ISR) and final (FSR) state.
The branchings in the shower are taken to be of lower virtualities than the hard interaction
introducing a virtuality (or time) ordering in the event.
This approach gives rise to several heavy quark production mechanisms, which we will
call \textit{pair creation}, \textit{flavour excitation} and \textit{gluon splitting}.
The names may be somewhat misleading since all three classes create pairs at
$\g \to \Q\Qbar$ vertices, but it is in line with the colloquial nomenclature.
The three classes are characterized as follows.
\begin{description}
\item [Pair creation] The hard subprocess is one of the two LO
parton fusion processes $\g\g \to \Q\Qbar$ or $\q\qbar \to \Q\Qbar$.
Parton showers do not modify the production cross sections, but only shift kinematics. 
For instance, in the LO description, the $\Q$ and $\Qbar$ have to emerge 
back-to-back in azimuth in order to conserve momentum, while the parton shower 
allows a net recoil to be taken by one or several further partons. 
\item [Flavour excitation] A heavy flavour from the parton distribution
of one beam particle is put on mass shell by scattering against a parton of 
the other beam, i.e. $\Q\q \to \Q\q$ or $\Q\g \to \Q\g$. When the $\Q$ is not a
valence flavour, it must come from a branching $\g \to \Q\Qbar$ of the
parton-distribution evolution. In most current sets of parton-distribution
functions, heavy-flavour distributions are assumed to vanish for
virtuality scales $Q^2 < m_{\Q}^2$. The hard scattering must therefore 
have a virtuality above $m_{\Q}^2$. When the initial-state shower is 
reconstructed backwards \cite{Sjostrand:PLB157}, the $\g \to \Q\Qbar$ branching
will be encountered, provided that $Q_0$, the lower cutoff of the shower,
obeys $Q_0^2 < m_{\Q}^2$. Effectively the processes therefore become
at least $\g\q \to \Q\Qbar\q$ or $\g\g \to \Q\Qbar\g$, with the possibility 
of further emissions. In principle, such final states could also be obtained 
in the above pair-creation case, but the requirement that the hard scattering
must be more virtual than the showers avoids double counting.
\item [Gluon splitting] A $\g \to \Q\Qbar$ branching occurs in the 
initial- or final-state shower but no heavy flavours are produced in the hard 
scattering. Here the dominant $\Q\Qbar$ source is gluons in 
the final-state showers since time-like gluons emitted in the initial state 
are restricted to a smaller maximum virtuality. Except at high energies, 
most initial state gluon splittings instead result in flavour excitation, already covered above.
An ambiguity of terminology exists with initial-state evolution chains where
a gluon first branches to $\Q\Qbar$ and the $\Q$ later emits another gluon
that enters the hard scattering. From 
an ideological point of view, this is flavour excitation, since it is related 
to the evolution of the heavy-flavour parton distribution. From a practical 
point of view, however, we choose to classify it as gluon splitting, 
since the hard scattering does not contain any heavy flavours. 
\end{description}

In summary, the three classes above are then characterized by having 2, 1 or 
0, respectively, heavy flavours in the final state of the LO hard subprocess.
Another way to proceed is to add next-to-leading order (NLO) perturbative 
processes, i.e the $\mathcal{O}(\alphas^3)$ corrections to the
parton fusion \cite{NDE2,Beenakker2}.
However, with our currently available set of calculational tools, the NLO 
approach is not so well suited for exclusive Monte Carlo studies where
hadronization is added to the partonic picture.

Flavour excitation and gluon splitting give significant contributions to the
total b cross section at LHC energies and thus must be considered when this is of
interest, see the following. However, NLO calculations probably do a better job
on the total b cross section itself (while, for the lighter c quark, production in
parton showers is so large that the NLO cross sections are more questionable).
The shapes of single heavy quark spectra are not altered as much as the correlations
between $\Q$ and $\Qbar$ when extra production channels are added.
Similar observations have been made when
comparing NLO to LO calculations \cite{Nason_and_CO}. Likewise, asymmetries between
single heavy quarks are also not changed much by adding further production channels,
so for simplicity we consider only the pair creation process here.

After an event has been generated at the parton level we add fragmentation to
obtain a hadronic final state. We use the Lund string fragmentation model.
Its effects on charm production were described in \cite{Norrbin:PLB442}.
Here we only summarize the main points.

In the string model, confinement is implemented by spanning strings between 
the outgoing partons. These strings correspond to a Lorentz-invariant 
description of a linear confinement potential with string tension 
$\kappa \approx 1$~GeV/fm. Each string piece has a colour charge at one end 
and its anticolour at the other. The double colour charge of the gluon 
corresponds to it being attached to two string pieces, while a quark is 
only attached to one. A diquark is considered as being in a colour 
antitriplet representation, and thus behaves (in this respect) like an 
antiquark. Then each string contains a colour triplet endpoint, a number 
(possibly zero) of intermediate gluons and a colour antitriplet end. An event
will normally contain several separate strings, especially at high energies
where $\g \to \q\qbar$ splittings occur frequently in the parton shower.

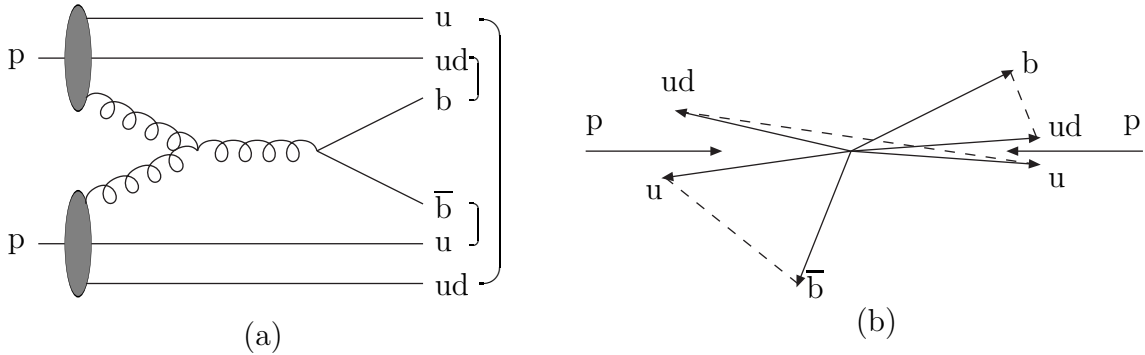
\begin{figure}[t]
\begin{picture}(210,130)(-10,-10)
\Text(100,-5)[]{(a)}
\Text(10,30)[r]{$\p$}
\Text(10,100)[r]{$\p$}
\Line(15,30)(25,30)
\Line(15,100)(25,100)
\GOval(30,30)(20,5)(0){0.5}
\GOval(30,100)(20,5)(0){0.5}
\Gluon(33,45)(75,65){4}{4}
\Gluon(33,85)(75,65){4}{4}
\Gluon(75,65)(120,65){4}{4}
\Line(120,65)(160,45)
\Line(120,65)(160,85)
\Text(165,45)[l]{$\bbar$}
\Text(165,85)[l]{$\b$}
\Line(33,15)(160,15)
\Line(33,115)(160,115)
\Text(165,15)[l]{$\u\d$}
\Text(165,115)[l]{$\u$}
\Line(35,30)(160,30)
\Line(35,100)(160,100)
\Text(165,30)[l]{$\u$}
\Text(165,100)[l]{$\u\d$}
\put(178,37.5){\oval(7,16)[r]}
\put(182,65){\oval(15,100)[r]}
\put(178,92.5){\oval(7,16)[r]}
\end{picture}
\hspace{5mm}
\begin{picture}(210,130)(0,-15)
\Text(110,-5)[]{(b)}
\LongArrow(0,60)(50,60)
\Text(0,70)[l]{$\p$}
\LongArrow(210,60)(160,60)
\Text(210,70)[r]{$\p$}
\LongArrow(100,60)(170,65)
\Text(175,70)[l]{$\u\d$}
\LongArrow(100,60)(160,90)
\Text(165,94)[l]{$\b$}
\LongArrow(100,60)(30,50)
\Text(26,47)[t]{$\u$}
\LongArrow(100,60)(80,10)
\Text(90,10)[r]{$\bbar$}
\LongArrow(100,60)(35,75)
\Text(35,87)[]{$\u\d$}
\LongArrow(100,60)(170,55)
\Text(175,49)[l]{$\u$}
\DashLine(170,65)(160,90){4}
\DashLine(30,50)(80,10){4}
\DashLine(35,75)(170,55){4}
\end{picture}\\[2mm]
\caption{Example of a string configuration in a $\p\p$ collision.
(a) Graph of the process, with brackets denoting the final colour singlet
subsystems. (b) Corresponding momentum space picture, with dashed lines
denoting the strings.
\label{asym:colourflow}}
\end{figure}

The string topology can be derived from the colour flow of the
hard process with some ambiguity arising from colour-suppressed terms.
Consider e.g. the LO process $\g\g \to \b\bbar$ where
two distinct colour topologies are possible. Representing the 
proton remnant by a $\u$ quark and a $\u\d$ diquark (alternatively $\d$ 
plus $\u\u$), one possibility is to have the three strings $\b$--$\u\d$, 
$\bbar$--$\u$ and $\u$--$\u\d$, Fig.~\ref{asym:colourflow}, and the other
is identical except the $\b$ is instead connected to the $\u\d$ diquark of
the other proton because the initial state is symmetric.

Once the string topology has been determined, the Lund string 
fragmentation model \cite{AGIS:PRP97} can be applied to describe the
nonperturbative hadronization. To first approximation, we assume that the 
hadronization of each colour singlet subsystem, i.e. string, can be considered 
separately from that of all the other subsystems. Presupposing that the 
fragmentation mechanism is universal, i.e. process-independent,
the good description of $\e^+\e^-$ annihilation data should carry over.
The main difference between $\e^+\e^-$ and hadron--hadron events is that 
the latter contain beam remnants which are colour-connected with the
hard-scattering partons.

Depending on the invariant mass of a string, practical considerations
lead us to distinguish the following three hadronization prescriptions:
\begin{description}
\item [Normal string fragmentation] 
In the ideal situation, each string has a large invariant mass. Then the 
standard iterative fragmentation scheme, for which the assumption of a 
continuum of phase-space states is essential, works well. The average 
multiplicity of hadrons produced from a string increases linearly with
the string `length', which means 
logarithmically with the string mass. In practice, this approach can be
used for all strings above some cutoff mass of a few GeV. 
\item [Cluster decay]
If a string is produced with a small invariant mass, perhaps only a single two-body
final state is kinematically accessible. In this case the standard iterative
Lund scheme is not applicable. We call such a low-mass string a cluster and consider
its decay separately. When kinematically possible, a $\Q$--$\qbar$ 
cluster will decay into one heavy and one light hadron by the production 
of a light $\q\qbar$ pair in the colour force field between the 
two cluster endpoints with the new quark flavour selected according to 
the same rules as in normal string fragmentation. The $\qbar$ cluster end 
or the new $\q\qbar$ pair may also  denote a diquark. In the latest version
of \mbox{\Py}, anisotropic decay of a cluster has been introduced, where the mass
dependence of the anisotropy has been matched to string fragmentation.
\item [Cluster collapse]
This is the extreme case of cluster decay, where the string 
mass is so small that the cluster cannot decay into two hadrons.
It is then assumed to collapse directly into a single hadron which
inherits the flavour contents of the string endpoints. The original 
continuum of string/cluster masses is replaced by a discrete set
of hadron masses, mainly $\mathrm{B}$ and $\mathrm{B}^*$ (or the corresponding 
baryon states). This mechanism plays a special r\^ole since it allows 
flavour asymmetries favouring hadron species that can inherit 
some of the beam-remnant flavour contents. Energy and momentum is not conserved
in the collapse so that some energy-momentum has to be taken from, or transferred
to, the rest of the event. In the new version, a scheme has been introduced
where energy and momentum are shuffled locally in an event.
\end{description}

We assume that the nonperturbative hadronization process does not change the 
perturbatively calculated total rate of bottom production. By local duality 
arguments \cite{Bloom:PRD4}, we further presume that the rate of cluster collapse
can be obtained from the calculated rate of low-mass strings. In the process
$\e^+\e^- \to \c\cbar$ local duality suggests that the sum of the
$\mathrm{J}/\psi$ and $\psi'$ cross sections approximately equal 
the perturbative $\c\cbar$ production cross section in the mass interval
below the $\D\Dbar$-threshold. Similar
arguments have also been proposed for $\tau$ decay to hadrons
\cite{BNP:NPB373} and shown to be accurate.
In the current case, the presence of other strings in the 
event also allows soft-gluon exchanges to modify
parton momenta as required to obtain the correct hadron masses.
Traditional factorization of short- and long-distance physics would 
then also preserve the total bottom cross section. Local duality and factorization,
however, do not specify \textit{how} to conserve the overall energy and 
momentum of an event when a continuum of $\bbar\d$ masses is to be replaced 
by a discrete $\mathrm{B}^0$. In practice, however, the different possible
hadronization mechanisms do not affect asymmetries much. The fraction of the string-mass
distribution below the two particle threshold effectively determines the total
rate of cluster collapse and therefore the asymmetry.

The cluster collapse rate depends on several model parameters. The most
important ones are listed here with the \mbox{\Py}~parameter values that we have used.
The \mbox{\Py}~parameters are included in the new default parameter set in \Py~\texttt{6.135}
and later versions.
\begin{itemize}
\item {\bf Quark masses}
The quark masses affect the threshold of the
string-mass distribution. Changing the quark mass shifts the string-mass
threshold relative to the fixed mass of the lightest two-body hadronic
final state of the cluster. Smaller quark
masses imply larger below-threshold production and an increased asymmetry.
The new default masses are \texttt{PMAS(1)}$=m_\u=$ \texttt{PMAS(2)}$=m_\d=$ \texttt{0.33D0},
\texttt{PMAS(3)}$=m_\s=$ \texttt{0.5D0}, \texttt{PMAS(4)}$=m_\c=$ \texttt{1.5D0}
and \texttt{PMAS(5)}$=m_\b=$ \texttt{4.8D0}.
\item {\bf Width of the primordial $\kt$ distribution.}
If the incoming partons are given small $\pt$ kicks in the initial state,
asymmetries can appear at larger $\pt$ since the beam remnants are given compensating
$\pt$ kicks, thus allowing collapses at larger $\pt$.
The new parameters are \texttt{PARP(91)=1.D0} and \texttt{PARP(93)=5.D0}.
\item {\bf Beam remnant distribution functions (BRDF).}
When a gluon is picked out of
the proton, the rest of the proton forms a beam remnant consisting, to first
approximation, of a quark and a diquark. How the remaining energy
and momentum should be split between these two is not known from first
principles. We therefore use different parameterizations of the splitting function and
check the resulting variations. We find significant differences only at large rapidities where
an uneven energy-momentum splitting tend to shift bottom quarks connected
to a beam remnant diquark more in the direction of the beam remnant, hence giving rise
to asymmetries at very large rapidities. We use an intermediate scenario in this
study, given by \texttt{MSTP(92)=3}.
\item {\bf Threshold behaviour between cluster decay and collapse.} Consider a
$\b\dbar$ cluster with an invariant mass at, or slightly above, the two particle
threshold. Should this cluster decay to two hadrons or collapse into one?
In one extreme point of view, a $\mathrm{B}\pi$ pair should always be formed 
when above this threshold, and never a single $\mathrm{B}$. In another extreme, the
two-body fraction would gradually increase at a succession of thresholds:
$\mathrm{B}\pi$, $\mathrm{B}^*\pi$, $\mathrm{B}\rho$, $\mathrm{B}^*\rho$, etc.,
where the relative probability for each channel is given by the standard 
flavour and spin mixture in string fragmentation.
In our current default model, we have chosen to steer a middle course
by allowing two attempts (\texttt{MSTJ(17)=2}) to find a possible pair of hadrons.
Thus a fraction  of events may collapse to a single resonance also above the
$\mathrm{B}\pi$ threshold, but $\mathrm{B}\pi$ is effectively weighted up. If a large
number of attempts had been allowed (this can be varied using the free parameter
\texttt{MSTJ(17)}), collapse would only become possible for cluster masses below the
$\mathrm{B}\pi$ threshold.
\end{itemize}

\begin{figure}
\begin{center}
\mbox{\epsfig{file=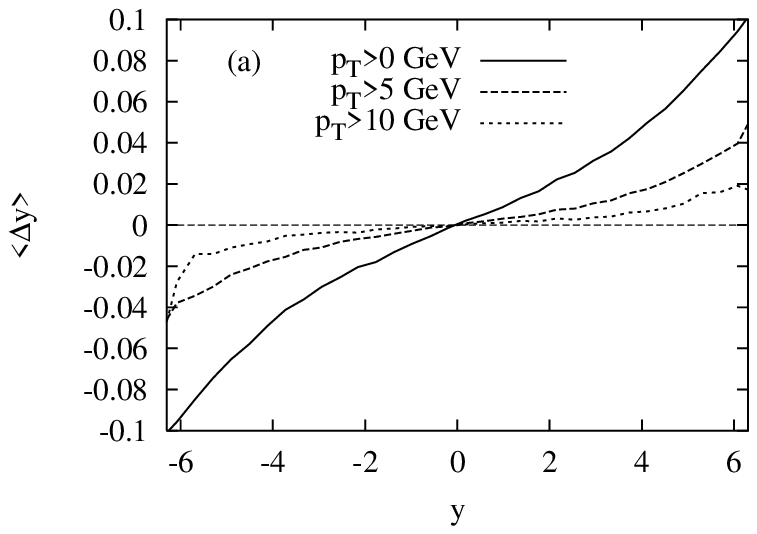}}
\mbox{\epsfig{file=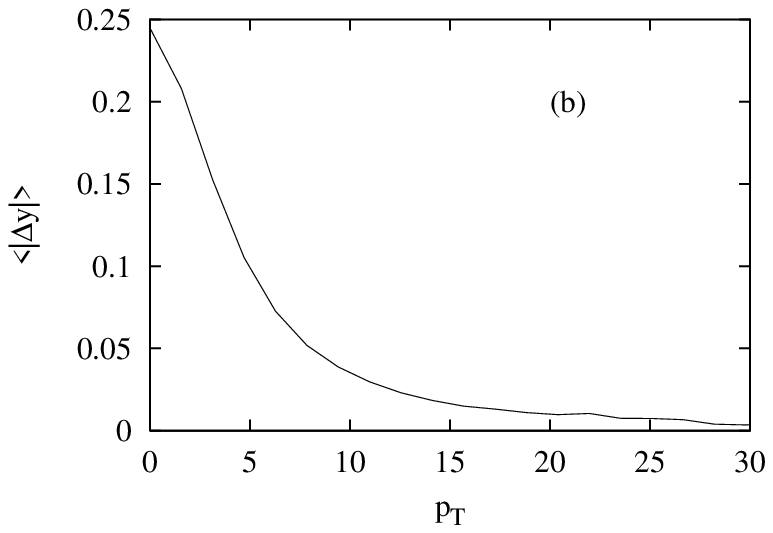}}
\end{center}
\caption{(a) Average rapidity shift $\Delta y = \langle y_\mathrm{B} - y_\mathrm{b} \rangle$
as a function of $y$ for some different $\pt$ cuts. (b) Average rapidity shift
$\langle |\Delta y| \rangle$ in the direction of ``the other end of the string''
that the bottom quark is connected to, i.e. ignoring the sign of the shift.}
\label{asym:drageffect}
\end{figure}

The colour connection between the produced heavy quarks and the beam remnants in the
string model gives rise to an effect called beam remnant drag. In an independent
fragmentation scenario the light cone energy momentum of the quark
is simply scaled by some factor picked from a fragmentation function.
Thus, on average the rapidity is conserved in the fragmentation process.
This is not necessarily so in string fragmentation, where
both string ends contribute to the four-momentum of the produced
heavy hadron. If the other end of the string is a beam remnant,
the hadron will be shifted in rapidity in the direction of the beam remnant
resulting in an increase in $|y|$. This beam-drag is shown
qualitatively in Fig.~\ref{asym:drageffect}, where the rapidity shift is shown as a
function of rapidity and transverse momentum. This shift is not directly accessible
experimentally, only indirectly as a discrepancy between the shape of perturbatively
calculated quark distributions and the data.

\section{Intrinsic Heavy Quarks}
\label{Asym:Intrinsic}
 
The wavefunction of a hadron in QCD can be represented as a
superposition of Fock state fluctuations, e.g. $\vert n_\V
\rangle$, $\vert n_\V \g \rangle$, $\vert n_\V \Q \Qbar \rangle$,
\ldots components where $n_\V \equiv \u\u\d$ for a proton.
When the projectile scatters in the target, the
coherence of the Fock components is broken and the fluctuations can
hadronize either by uncorrelated fragmentation as for leading twist
production or coalescence with
spectator quarks in the wavefunction \cite{Brodsky:PLB93,Brodsky:NPB369}.
The intrinsic heavy quark Fock components are generated by virtual
interactions such as $\g \g \rightarrow \Q \Qbar$ where the
gluons couple to two or more projectile valence quarks. Intrinsic
$\Q\Qbar$ Fock states are dominated by configurations with
equal rapidity constituents so that, unlike sea quarks generated
from a single parton, the intrinsic heavy quarks carry a large
fraction of the parent momentum \cite{Brodsky:PLB93}.
 
The frame-independent probability distribution of an $n$--particle
$\b \bbar$ Fock state is
\begin{eqnarray}
\frac{dP^n_{\rm ib}}{dx_i \cdots dx_n} = N_n 
\frac{\delta(1-\sum_{i=1}^n x_i)}{(m_h^2 - \sum_{i=1}^n
(\widehat{m}_i^2/x_i) )^2} \, \, ,
\label{icdenom}
\end{eqnarray}
where $\widehat{m}_i^2 =k^2_{\perp,i}+m^2_i$ is the effective transverse mass
of the $i^{\rm th}$ particle and $x_i$ is the light-cone momentum
fraction.   The probability, $P^n_{\rm ib}$, is normalized by $N_n$
and $n=5$ for baryon production from the $|n_\V \b\bbar \rangle$ 
configuration.  The delta function
conserves longitudinal momentum.  The dominant Fock configurations
are closest to the light-cone energy shell and therefore the invariant
mass, $M^2 = \sum_i \widehat{m}_i^2/ x_i$, is minimized. 
Assuming $\langle \vec k_{\perp, i}^2 \rangle$ is
proportional to the square of the constituent quark mass, we choose
$\widehat{m}_\q = 0.45$ GeV, $\widehat{m}_\s =
0.71$ GeV, and $\widehat{m}_\b = 5$ GeV \cite{VBH:NPB360,VBH:NPB383}.
 
The $\xF$ distribution for a single bottom hadron produced from an
$n$-particle intrinsic bottom state can be related to
$P^n_{\rm ib}$ and the inelastic $\p\p$ cross section by
\begin{eqnarray}
\frac{\sigma^H_{\rm ib}(\p\p)}{d\xF} = \frac{dP_H}{d\xF} \sigma_{\p\p}^{\rm in}
\frac{\mu^2}{4 \widehat{m}_\b^2}  \alpha_s^4(M_{\b\bbar}) \, \, .
\label{icsign}
\end{eqnarray}
The probability distribution is the sum of all contributions from the $|n_\V \b
\bbar \rangle$ and the $|n_\V \b
\bbar \q \qbar \rangle$ configurations with $\q = \u$, $\d$, and $\s$
and includes uncorrelated fragmentation and coalescence, as described below, 
when appropriate \cite{Vogt:LBNL43095}.
The factor of $\mu^2/4 \widehat{m}_\b^2$ arises from the soft
interaction which breaks the coherence of the Fock state. 
We take  $\mu^2 \sim 0.1$
GeV$^2$ \cite{GV:B539}. The intrinsic charm probability, 
$P^5_{\rm ic} = 0.31$\%, was determined from analyses of the EMC
charm structure function data \cite{EMC:PLB110}. The intrinsic bottom
probability is scaled from the intrinsic charm probability by the square of the
transverse masses,  $P_{\rm ib} = P_{\rm ic} (\widehat{m}_\c/\widehat{m}_\b)^2$.
The intrinsic bottom cross
section is reduced relative to the intrinsic charm cross section by a factor of
$\alphas^4(M_{\b\bbar})/\alphas^4(M_{\c\cbar})$ \cite{VB:NPB438}.
Taking these factors into account, we obtain $\sigma^5_{\rm ib}(p N) \approx 7$ nb at 14 TeV.  

There are two ways of producing bottom hadrons from intrinsic $\b\bbar$ states.
The first is by uncorrelated fragmentation.  
If we assume that the $\b$ quark fragments into a $\mathrm{B}$ meson, the $\mathrm{B}$
distribution is
\begin{eqnarray}
\frac{d P^{nF}_{\rm ib}}{dx_{\mathrm{B}}} = \int dz \prod_{i=1}^n dx_i
\frac{dP^n_{\rm ib}}{dx_1 \ldots dx_n} \frac{D_{B/b}(z)}{z} \delta(x_{\mathrm{B}} - z
x_\b) \, \, ,
\label{icfrag}
\end{eqnarray}
These distributions are assumed for all intrinsic bottom production by uncorrelated
fragmentation with $D_{H/b}(z) = \delta(z-1)$.  At low $\pt$, this
approximation should not be too bad, as seen in fixed target production
\cite{VBH:NPB383}.

If the projectile has the corresponding valence quarks, the bottom quark can also hadronize
by coalescence with the valence spectators. The coalescence distributions are specific for
the individual bottom hadrons. It is reasonable to assume that the intrinsic
bottom Fock states are fragile and can easily materialize into bottom hadrons
in high-energy, low momentum transfer reactions through coalescence.   
The coalescence contribution to bottom hadron production is
\begin{eqnarray}
\frac{d P^{nC}_{\rm ib}}{dx_H} = \int \prod_{i=1}^n dx_i
\frac{dP^n_{\rm ib}}{dx_1 \ldots dx_n} \delta(x_H - x_{H_1}-\cdots - 
x_{H_{n_\V}}) \, \, . 
\label{iccoalD}
\end{eqnarray}
where the 
coalescence function is simply a delta function combining the momentum
fractions of the quarks in the Fock state configuration that make up the
valence quarks of the final-state hadron.

Not all bottom hadrons can be produced from the minimal intrinsic bottom
Fock state configuration, $|n_\V \b\bbar \rangle$. 
However, coalescence can also occur within higher fluctuations of the
intrinsic bottom Fock state.  For example, in the proton, the $\mathrm{B}^-$ and
$\Xi_\b^0$ can be produced by coalescence from $|n_\V \b
\bbar\u\ubar \rangle$ and $|n_\V \b\bbar\s\sbar \rangle$ configurations.
These higher Fock state probabilities can be obtained using earlier results
on $\psi \psi$ pair production \cite{VB:NPB478,VB:PLB349}. 
If all the measured $\psi \psi$ pairs
\cite{Badpsipsi} arise from $|n_\V \c \cbar\c\cbar \rangle$ 
configurations, $P_{\rm icc} \approx 4.4\%\ P_{\rm
ic}$ \cite{VB:PLB349,RV:NPB446}. It was found that the probability of a
$|n_\V \c\cbar\q\qbar \rangle$
state was then $P_{\rm icq} = (\hat m_\c/\hat m_\q)^2 P_{\rm icc}$ \cite{VB:NPB478}.  
If we then assume $P_{\rm ibq} = (\hat m_\c/\hat m_\b)^2
P_{\rm icq}$, we find that 
\begin{eqnarray}
P_{\rm ibq} \approx \left( \frac{\widehat{m}_\c}{\widehat{m}_\b}
\right)^2 \left( \frac{\widehat{m}_\c}{\widehat{m}_\q}
\right)^2 P_{\rm icc} \, \, ,
\label{icrat}
\end{eqnarray}
leading to $P_{\rm ibu} = P_{\rm ibd} \approx 70.4\%\ P_{\rm ib}$
and $P_{\rm ibs} \approx 28.5\%\ P_{\rm ib}$.  To go to still higher
configurations, one can make similar assumptions. However, as more
partons are included in the Fock state, the coalescence
distributions soften and approach the fragmentation distributions,
eventually producing bottom hadrons with less momentum
than uncorrelated fragmentation from the minimal $\b\bbar$
state if a sufficient number of $\q\qbar$ pairs are included.
There is then no longer any advantage to introducing more light
quark pairs into the configuration---the relative probability will
decrease while the potential gain in momentum is not significant.
Therefore, we consider
production by fragmentation and coalescence from the minimal state and the next
higher states with $\u\ubar$, $\d\dbar$ and $\s\sbar$ pairs.
 
The probability distributions entering Eq.~(\ref{icsign}) for $\mathrm{B}^0$ and 
$\Bbar^0$ are
\begin{eqnarray} 
\frac{dP_{\mathrm{B}^0}}{d\xF} & = & \frac{1}{2} \left( \frac{1}{10} \frac{dP_{\rm
ib}^{5F}}{d\xF} + \frac{1}{4} \frac{dP_{\rm ib}^{5C}}{d\xF} \right) + 
 \frac{1}{2} \left( \frac{1}{10} \frac{dP_{\rm
ibu}^{7F}}{d\xF} + \frac{1}{5} \frac{dP_{\rm ibu}^{7C}}{d\xF} \right) \nonumber
\\ &   & + \, \frac{1}{2} \left( \frac{1}{10} \frac{dP_{\rm
ibd}^{7F}}{d\xF} + \frac{2}{5} \frac{dP_{\rm ibd}^{7C}}{d\xF} \right) +
 \frac{1}{2} \left( \frac{1}{10} \frac{dP_{\rm
ibs}^{7F}}{d\xF} + \frac{1}{5} \frac{dP_{\rm ibs}^{7C}}{d\xF} \right) \\
 \frac{dP_{\Bbar^0}}{d\xF} & = & 
\frac{1}{10} \frac{dP_{\rm ib}^{5F}}{d\xF} +
 \frac{1}{10} \frac{dP_{\rm ibu}^{7F}}{d\xF} + \frac{1}{2} \left( \frac{1}{10}
\frac{dP_{\rm ibd}^{7F}}{d\xF} + \frac{1}{8} \frac{dP_{\rm ibd}^{7C}}{d\xF} 
\right) + \frac{1}{10} \frac{dP_{\rm ibs}^{7F}}{d\xF} \, \, .
\end{eqnarray}
See Ref.~\cite{Vogt:LBNL43095} for more details and the probability distributions of
other bottom hadrons.

\section{Model predictions}
\begin{figure}
\begin{center}
\mbox{\epsfig{file=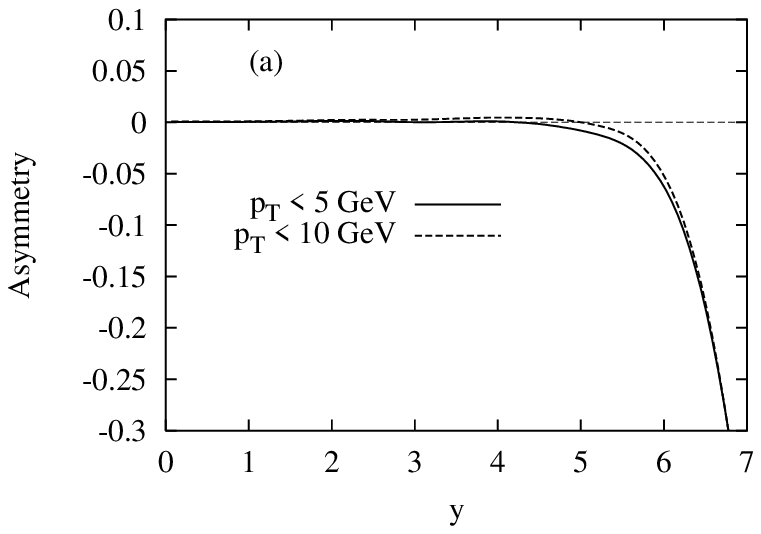}}
\mbox{\epsfig{file=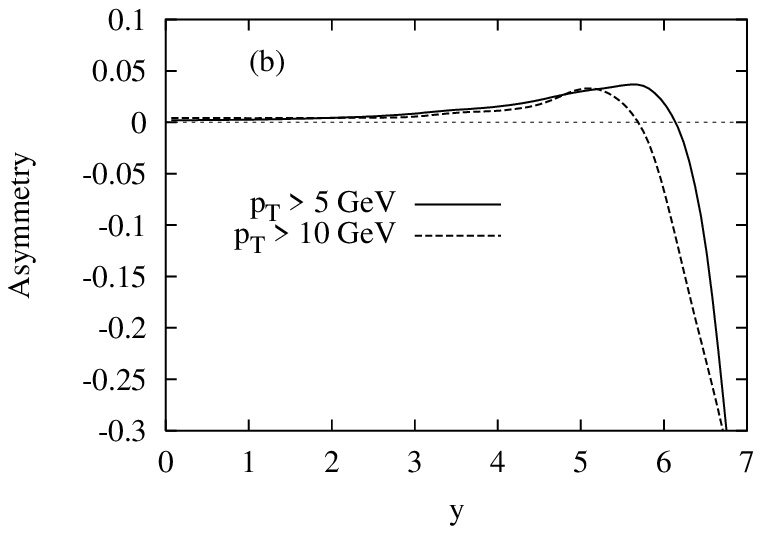}}
\end{center}
\caption{The asymmetry, $A=\frac{\sigma(\mathrm{B}^0) - \sigma(\Bbar^0)}
{\sigma(\mathrm{B}^0) + \sigma(\Bbar^0)}$, as a function of rapidity for
different $\pt$ cuts: (a) $\pt < 5,10$ GeV and (b) $\pt > 5, 10$ GeV using
parameter set 1 as described in the text.}
\label{asym:string_asym}
\end{figure}

In this section we present some results from both models. Fig.~\ref{asym:string_asym}
shows the asymmetry between $\mathrm{B}^0$ and $\Bbar^0$ as a function of $y$ for several
$\pt$ cuts in the string model. The asymmetry
is essentially zero for central rapidities and increases slowly with rapidity.
When the kinematical limit is approached, the asymmetry changes sign for small
$\pt$ because of the drag-effect since $\b$-quarks are often connected to diquarks
from the proton beam remnant, Fig.~\ref{asym:colourflow}, thus producing $\Bbar^0$
hadrons which are shifted
more in rapidity than $\mathrm{B}^0$. Cluster collapse, on the other hand, tend to
enhance the production of leading particles (in this case $\mathrm{B}^0$) so
the two mechanisms give rise to asymmetries with different signs. Collapse is
the main effect at small rapidities while eventually at very large $y$, the drag effect
dominates.

In Table~\ref{asym:parameterdep} we study the parameter dependence of the asymmetry by
looking at the integrated asymmetry for different kinematical regions using three
different parameter sets:
\begin{itemize}
\item {\bf Set 1} is the new default as presented in Section~\ref{Asym:Lundstring}.
\item {\bf Set 2} The same as Set 1 except it uses simple counting rules in the beam remnant
splitting, i.e. each quark get on average one third of the beam remnant energy-momentum.
\item {\bf Set 3} The old parameter set, before fitting to fixed-target data, is included as a
reference. This set is characterized by current algebra masses, lower intrinsic $\kt$,
and an uneven sharing of beam remnant energy-momentum.
\end{itemize}

We see that in the central region the asymmetry is generally very small whereas for
forward (but not extremely forward) rapidities and moderate $\pt$ the asymmetry
is around 1--2\%. In the very forward region at small $\pt$, drag asymmetry dominates
which can be seen from the change in sign of the asymmetry. The asymmetry is fairly
stable under moderate variations in the parameters even though the difference between
the old and new parameter sets (Set 1 and 3) are large in the central region.
Set 1 typically gives rise to smaller asymmetries.

\begin{table}
\begin{center}
\begin{tabular}{|l|c|c|c|} \hline
Parameters	& $|y|<2.5$, $\pt>5$ GeV& $3<|y|<5$, $\pt>5$ GeV& $|y|>3$, $\pt<5$ GeV \\ \hline
Set 1		& 0.003(1)		& 0.015(2)		& $-$0.008(1) \\ \hline
Set 2		& $-$0.000(2)		& 0.009(3)		& $-$0.005(2) \\ \hline
Set 3		& 0.013(2)		& 0.020(3)		& $-$0.018(2) \\ \hline
\end{tabular}
\end{center}
\caption{Parameter dependence of the asymmetry in the string model. The statistical
error in the last digit is shown in parenthesis (95\% confidence).}
\label{asym:parameterdep}
\end{table}

\begin{figure}[htb]
\setlength{\epsfxsize=0.95\textwidth}
\setlength{\epsfysize=0.5\textheight}
\centerline{\epsffile{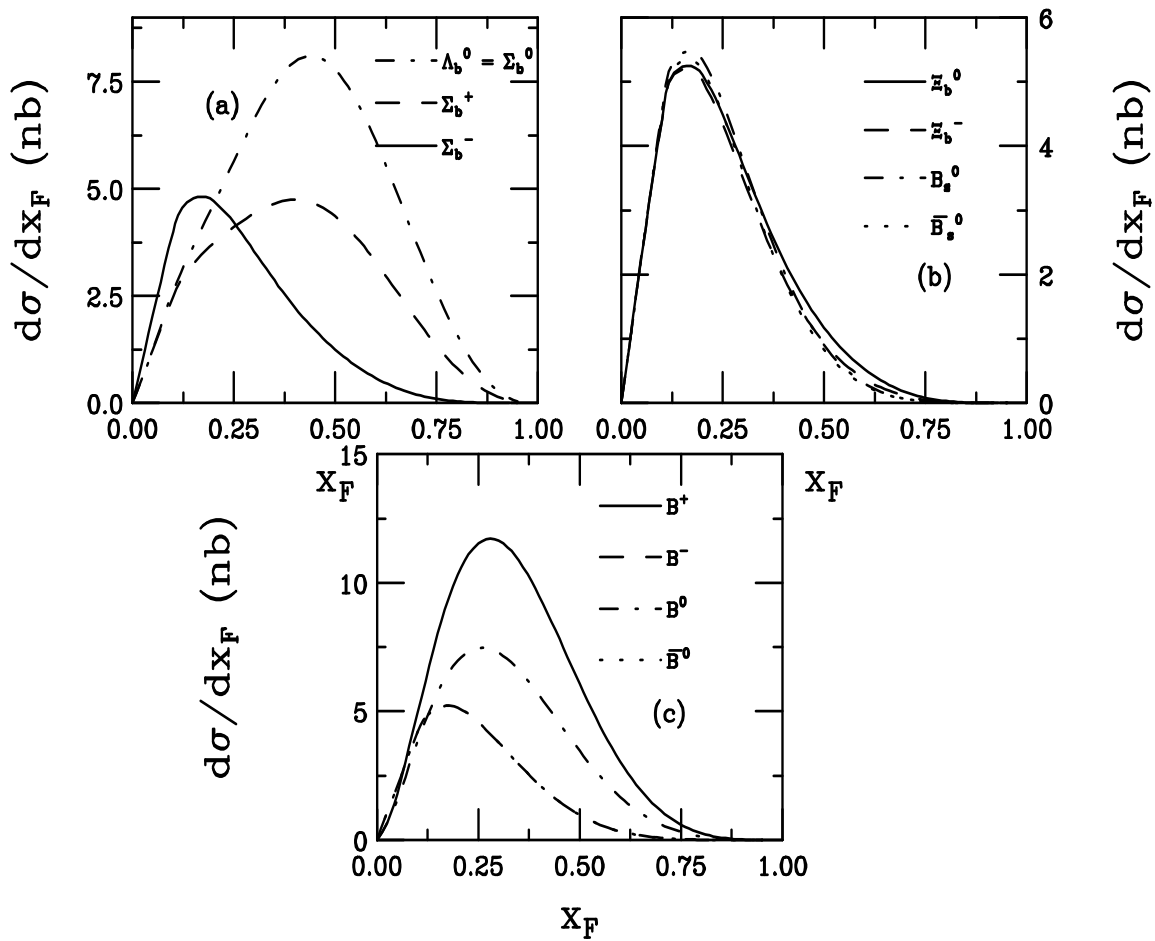}}
\caption[]{ Predictions for bottom hadron production are given for $\p\p$
collisions at 14 TeV.  The bottom baryon distributions are given in (a) for
$\Lambda_\b^0=\Sigma_\b^0$ (dot-dashed), $\Sigma_\b^+$ (dashed), and $\Sigma_\b^-$
(solid).  The bottom-strange distributions are shown in (b) for $\Xi_\b^0$
(solid), $\Xi_\b^-$ (dashed), $\mathrm{B}_\s^0$ (dot-dashed), and $\Bbar_s^0$
(dotted).  In (c), the $\mathrm{B}$ meson distributions are given: $\mathrm{B}^+$ (solid),
$\mathrm{B}^-$ (dashed), $\mathrm{B}^0$ (dot-dashed), and $\Bbar^0$ (dotted).
The $\mathrm{B}^-$ and $\Bbar^0$ distributions are virtually identical.}
\label{pro800}
\end{figure}

The cross sections for all intrinsic bottom hadrons are given as a function of $\xF$ in
Fig.~\ref{pro800}. The bottom baryon distributions are shown in
Fig.~\ref{pro800}(a).  The $\Lambda_\b^0$ ($\Sigma_\b^0$) distributions are the
largest and most forward peaked of all the distributions. The $\Sigma_\b^-$ 
is the smallest and the softest, similar to that of the
bottom-strange mesons and baryons shown in Fig.~\ref{pro800}(b).  The different
coalescence probabilities assumed for hadrons from the $|\u\u\d\b\bbar\s\sbar\rangle$
configuration have little real effect on the shape of the
cross section, dominated by independent fragmentation. Of the $\mathrm{B}$ mesons
shown in Fig.~\ref{pro800}(c), the $\mathrm{B}^+$ and $\mathrm{B}^0$ cross sections are the 
largest since both can be produced from the 5 particle configuration. The
$\mathrm{B}^-$ and $\Bbar^0$ distributions are virtually identical. We 
note that the $\xF$ distributions of other bottom hadrons
not included in the figure would be similar to the bottom-strange hadrons since
they would be produced by fragmentation only.

\begin{figure}[htb]
\setlength{\epsfxsize=0.95\textwidth}
\setlength{\epsfysize=0.25\textheight}
\centerline{\epsffile{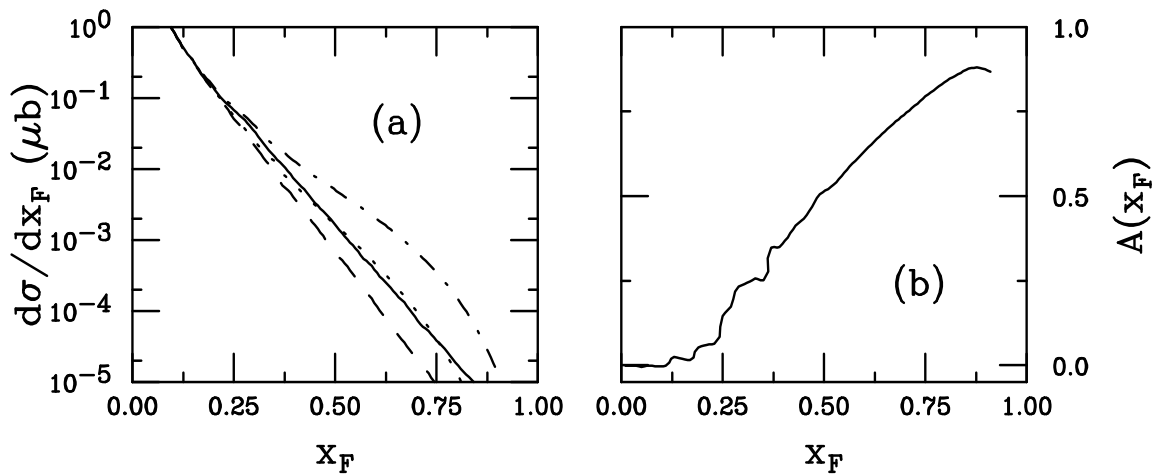}}
\caption[]{ (a) Leading-twist predictions for $\mathrm{B}^0$ (solid) and $\Bbar^0$
(dashed) using independent fragmentation.
Model predictions for $\mathrm{B}^0$ (dot-dashed) and $\Bbar^0$ (dotted)
distributions from Eq.~(\ref{tcmodel}). (b)
The asymmetry between $\mathrm{B}^0$ and $\Bbar^0$, the
dot-dashed and dotted curves in (a), is also given.}
\label{asymlhc}
\end{figure}

The $\xF$ distribution for final-state hadron $H$ is the sum
of the leading-twist fusion and intrinsic bottom components,
\begin{eqnarray}
\frac{d\sigma^H_{hN}}{d\xF} = \frac{d\sigma^H_{\rm lt}}{d\xF} +
\frac{d\sigma^H_{\rm ib}}{d\xF} \, \, .
\label{tcmodel}
\end{eqnarray}
The intrinsic bottom cross sections from Section~\ref{Asym:Intrinsic}
are combined with a leading twist calculation using independent
fragmentation where drag effects are not included.
The leading twist results have been smoothed and
extrapolated to large $\xF$ \footnotetext{Thanks
to J. Klay at UC Davis for extending the curves to large $\xF$.}
to facilitate a comparison with the intrinsic
bottom calculation.  The resulting total $\mathrm{B}^0$ and $\Bbar^0$
distributions are shown in Fig.~\ref{asymlhc}, along with the corresponding
asymmetry. Note that since the intrinsic heavy quark $\pt$ distributions are
more steeply falling than the leading twist, we only consider $\pt < 5$ GeV.
The distributions are drawn to emphasize the high $\xF$ region where the
distributions differ.  The asymmetry is $\sim 0.1$ at $\xF \sim 0.25$,
corresponding to $y\sim 6.5$.  Therefore, intrinsic bottom should not be a
significant source of asymmetries.

\section{Summary}
To summarize, we have studied possible production asymmetries between $\b$
and $\bbar$ hadrons, especially $\mathrm{B}^0$ and $\Bbar^0$, as predicted
by the Lund string fragmentation model and the intrinsic heavy quark model.
We find negligible asymmetries for central rapidities and large $\pt$
(in general, less than 1\%). For some especially favoured kinematical ranges such as
$y > 3$ and $5 < \pt < 10$ GeV the collapse asymmetry could be as high as 1--2\%. Intrinsic
bottom becomes important only for $\xF>0.25$ and $\pt<5$ GeV, corresponding to $y>6.5$.

\end{document}